\documentclass[twocolumn,twoside,slac]{revtex4}
\usepackage{graphicx}
\usepackage{fancyhdr}
\begin{document}

\title{LCIO - A persistency framework for linear collider simulation studies}


\author{F. Gaede}
\affiliation{DESY, 22607 Hamburg, Germany}

\author{T. Behnke}
\affiliation{DESY and SLAC}
\author{N. Graf, T. Johnson}
\affiliation{SLAC, Stanford, CA 94025, USA}

\begin{abstract}
Almost all groups involved in linear collider detector studies have their own 
simulation software framework. Using a common persistency scheme would allow 
to easily share results and compare reconstruction algorithms.
We present such a persistency framework, called LCIO (Linear Collider I/O).
The framework has to fulfill the requirements of the different groups 
today and be flexible enough to be adapted to future needs. 
To that end we define an `abstract object persistency layer' that will be 
used by the applications. A first implementation, based on a sequential file 
format (SIO) is completely separated from the interface, thus allowing to 
support additional formats if necessary.
The interface is defined with the AID (Abstract Interface Definition) tool 
from freehep.org that allows creation of Java and C++ code synchronously. 
In order to make use of legacy software a Fortran interface is also provided.
We present the design and implementation of LCIO.
\end{abstract}

\maketitle

\thispagestyle{fancy}


\section{INTRODUCTION \label{intro}}
\subsection{Motivation \label{sec_motivation}}
Most groups involved in the international linear collider detector studies
have their own simulation software applications. These applications are written 
in a variety of languages and support their own file formats for making data 
persistent in between the different stages of the simulation 
{\it(generator, simulation, reconstruction and analysis)}. 
The most common languages are Java and C++, but also 
Fortran is still being used.
In order to facilitate the sharing of results as well as comparing different 
reconstruction algorithms, a common persistency format would be desirable.
Such a common scheme could also serve as the basis for a common software framework to 
be used throughout the international simulation studies. The advantages of such a 
framework are obvious, primarily to avoid duplication of effort.

\begin{figure}
\includegraphics[width=80mm]{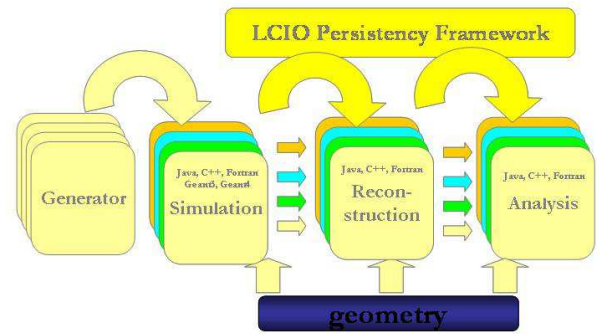}
\caption{Schematic overview of the simulation software chain. 
LCIO defines a persistency framework for the simulation and reconstruction stages. 
It is suited to replace the  plethora of different data and file formats currently 
being used in linear collider simulation studies (denoted by the small arrows).
\label{fig_motivation}}
\end{figure}

LCIO is a complete persistency framework that addresses these issues. 
To that end LCIO defines a data model, a user interface (API) and a file format for the 
simulation and reconstruction stages (fig.~\ref{fig_motivation}).

\subsection{Requirements \label{sec_requirements}}
In order to allow the integration of LCIO into currently existing software frameworks
a JAVA, a C++ and a Fortran version of the user interface are needed. The data model
defined by LCIO has to fulfill the needs of the ongoing simulation studies and at the 
same time be flexible for extensions in order to cope with future requirements.
It has become good practice to design persistency systems in a way that hides the
actual implementation of the data storage mechanism from the user code.
This allows to change the underlying data format without making nontrivial changes to 
existing code. In particular when writing software for an experiment that still is in the 
R\&D phase one has to anticipate possible future changes in the persistency 
implementation.

There are three major use cases for LCIO: writing data ({\em simulation}), reading 
and extending data ({\em reconstruction}) and read only access ({\em analysis}). 
These have to be reflected by the design of the user interface (API). 

One of the most important requirements is that an implementation is needed as soon as 
possible. The more software which has been written using incompatible persistency schemes, 
the more difficult it will be to incorporate a common approach.
By choosing a simple and lightweight design for LCIO we will be able to keep the 
development time to a minimum.  

% \begin{figure*}[t]
% \centering
% \includegraphics[width=135mm]{schematic_classes.eps}
% \caption{Example of full width figure.\label{JACpic2-f1}} 
% \end{figure*}

% \subsection{References}
% All bibliographical references should be numbered and listed at
% the end of the paper in a section called ``References.'' When
% referring to a reference in the text, place the corresponding
% reference number in square brackets~\cite{exampl-ref}.

\section{DATA MODEL \label{sec_datamodel}}

Before going into the details of the software design and realization of LCIO we need to
briefly describe the data model. One of the main goals of LCIO is the unification of 
software that is used in the international  linear collider community. Thus the data 
model has to fulfill the current needs of the different groups working in the field
as well as being expandable for future requirements.
\begin{figure}
\includegraphics[width=80mm]{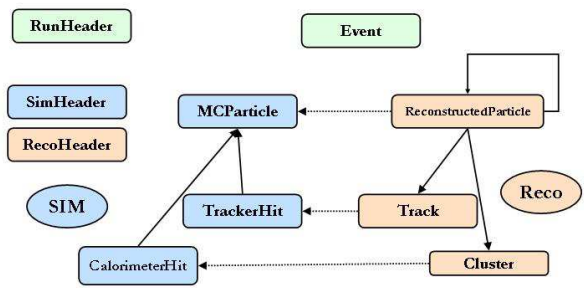}
\caption{Overview of the data model defined by LCIO. The boxes correspond to separate
data entities (classes) and the arrows denote relationships between these entities.
Not shown are extension objects for additional information that users might want to add.
\label{fig_datamodel}}
\end{figure}
Figure~\ref{fig_datamodel} shows the data entities that are defined by LCIO. 
The attributes currently stored for each entity basically form a superset of the
attributes used in two simulation software frameworks, namely the Java based 
{\em hep.lcd}~\cite{ref_heplcd} framework, used by some U.S. groups and the European
framework - consisting of the {\em Mokka}~\cite{ref_mokka} simulation and 
{\em Brahms}~\cite{ref_brahms} reconstruction program. 
These two frameworks are also the first implementors of LCIO. As other groups 
join in using LCIO,  their additional requirements can be taken into 
account by extending the data model as needed.
A detailed description of the attributes can be found at~\cite{ref_lcio}.

Various header records hold generic descriptions of the stored data. 
The {\em Event} entity holds collections of simulation and reconstruction output 
quantities.
{\em MCParticle} is  the list of generated particles, possibly 
extended by particles created during detector response simulation 
(e.g. $K_s \rightarrow \pi^+\pi^-$).
The simulation program adds hits from tracking detectors and calorimeters 
to the two generic hit entities {\em TrackerHit} and {\em CalorimeterHit}, thereby referencing the 
particles that contributed to the hit at hand. Pattern recognition and cluster algorithms
in the reconstruction program create tracks and clusters - stored in {\em Track} and 
{\em Cluster} respectively - in turn referencing the contributing hit objects. These 
references are optional as we foresee that hits may be dropped at some point in time.
So called 'particle flow' algorithms create the list of {\em ReconstructedParticles}
from tracks and clusters. This is generally done by taking the tracker measurement alone 
for charged particles and deriving the kinematics for neutral particles from clusters 
not assigned to tracks. By combining particles from the original list of reconstructed 
particles additional lists of the same type might be added to the event, (e.g. heavy 
mesons, jets, vertices).  Objects in these lists will point back to the constituent 
particles (self reference). 

\section{IMPLEMENTATION \label{sec_implementation}}

As mentioned in the previous section, LCIO will have a Java, C++ and a Fortran 
implementation. As it is good practice nowadays we choose an object oriented design for 
LCIO~(\ref{sec_design}). 
The visible user interfaces (API) for Java and C++ are kept as close as  
this is feasible with the two languages~(\ref{sec_javacpp}).
For Fortran we try to map as much as possible of the OO-design onto the
actual implementation~(\ref{sec_fortran}).

\subsection{Design \label{sec_design}}

\begin{figure*}
\includegraphics[height=100mm, width=150mm]{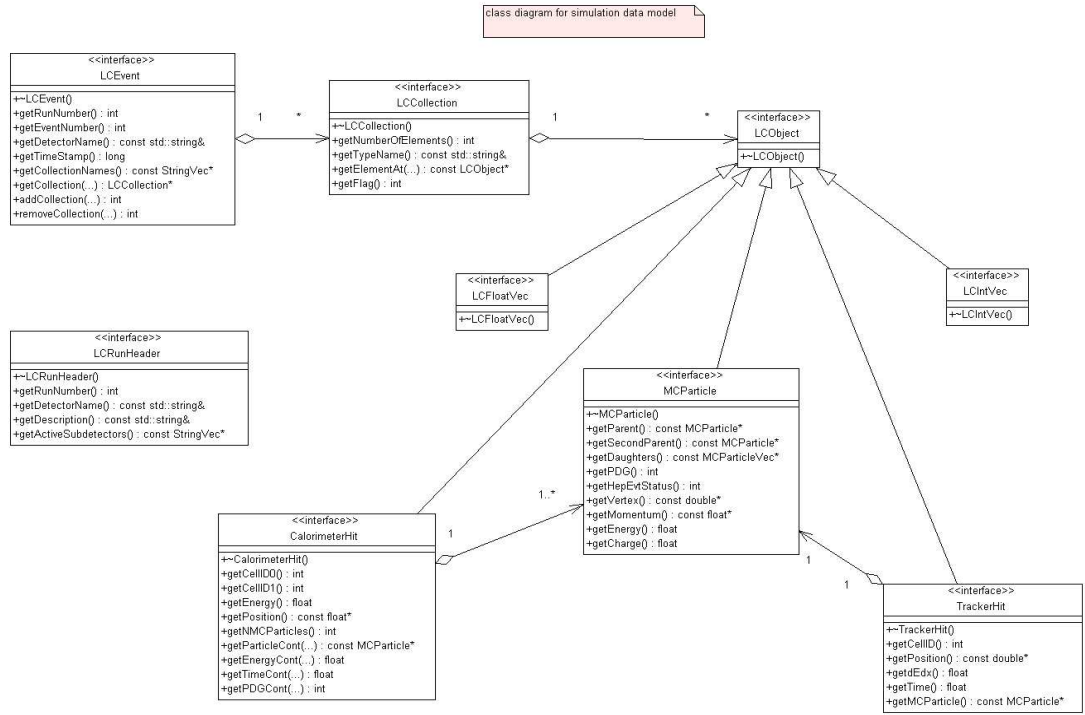}
\caption{UML class diagram showing the complete data model for simulation output. Event 
data is stored in LCEvent objects that in turn hold untyped collections (LCCollection).
The tagging interface LCObject is needed for C++ which doesn't have a common base class.
Existing simulation applications can use LCIO as an output format by implementing the 
shown interface within their existing classes. 
\label{fig_simclass}}
\end{figure*}

Here we give an overview of the basic class design in LCIO and how 
this reflects the requirements described in section~\ref{sec_requirements}.
One of the main use cases is to write data from a simulation
program and to read it back into a reconstruction program using LCIO. 
In order to facilitate the incorporation of LCIO for writing data into existing 
simulation applications we decided to have a minimal interface for the data objects. 
This minimal interface has as small a number of methods as possible, making it easy to 
implement within already existing classes. 
Figure~\ref{fig_simclass} shows
% Surround figure environment with turnpage environment for landscape
%\begin{turnpage}
%\end{turnpage}
a UML class diagram of this minimal interface. The LCIO implementation uses just this 
interface to write data from a simulation program. 
The {\em LCEvent} holds untyped collections of the data entities described in 
section~\ref{sec_datamodel}. In order to achieve this for C++ we introduced a 
tagging interface {\em LCObject}, that does not introduce any additional functionality.
{\em LCFloatVec} and {\em LCIntVec} enable the user to store arbitrary numbers of 
floating point or integer numbers per collection (e.g by storing a vector of size three
in a collection that runs parallel to a hit collection one could add a position to every 
hit from a particular subdetector, if this was needed for a particular study).

The same interface can be used for read only access to the data. 
For the case of reading and modifying data, e.g. in a reconstruction 
program, we provide default implementations for every object in the minimal 
interface. These implementation classes also have {\it set methods} and can be 
instantiated by the user for writing LCIO data files.
\begin{figure*}
\includegraphics[height=100mm, width=150mm]{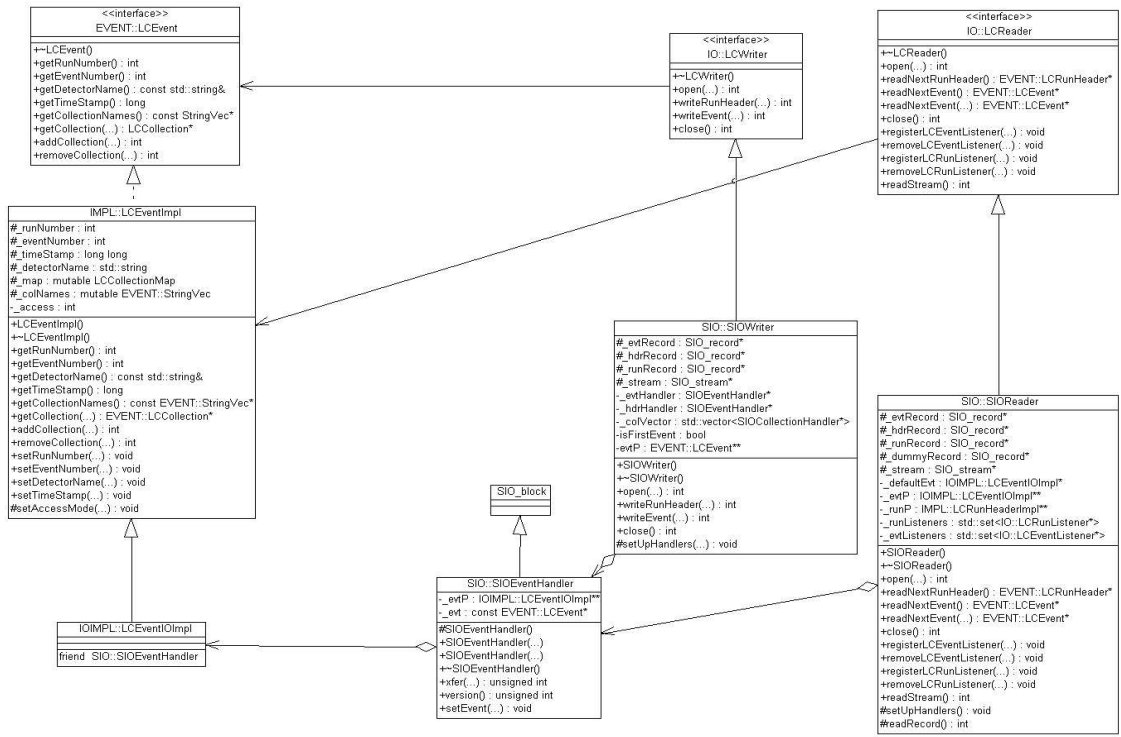}
\caption{UML class diagram showing the relationship between abstract user interface and 
implementation for the data entities and the IO-classes. Users only apply the abstract
interface ({\em top}) or instantiate the default implementation ({\em left}). The 
concrete classes for the IO ({\em lower right}) are hidden by a factory pattern.
\label{fig_ioimplclass}}
\end{figure*}
Figure~\ref{fig_ioimplclass} shows the implementation {\em LCEventImpl} of
the {\em LCEvent} interface together with the interface and implementation of the 
IO-classes. The {\em LCWriter} uses just the abstract interface of the data objects,
thus not relying on the default implementation. The {\em LCReader} on the  other hand 
uses the default implementations provided by LCIO. These are instantiated when reading 
events and accessed by the user either via the abstract interface, i.e. read only, 
or their concrete type for modification of the data.

The concrete implementations {\em SIOWriter} and {\em SIOReader}~(\ref{sec_sio})
are hidden from the user by exploiting a factory pattern. Therefore user code does not 
depend on the concrete data format. 
The interface of {\em LCWriter} is rather simple - it allows 
to open and close files (data streams) and to store event and run data.
The {\em LCReader} interface allows to access data in several ways:

\begin{itemize} 

\item{read run by run in the file\\ 
  - provides no access to event information }
\item{read event by event in the file\\
  - standard way of analyzing events in a simple analysis program}
\item{quasi direct access to a given event \\ 
  - currently implemented using a fast skip 
mechanism~(\ref{sec_sio})}
\item{ call-back mechanism to read run and event data \\
  - users can register modules that implement a listener interface (observer pattern) 
  with  the {\em LCReader} \\
  - these modules will be called in the order they have been 
  registered\\ 
  - this can be used e.g. in a reconstruction program where one typically has separate 
  modules for different algorithms that have to be called in a given order}
\end{itemize} 

When reading data, the minimal interface described above just provides access to the 
data in the form it had been made persistent. Sometimes it might be convenient to
have different parameterizations of the data. Convenience methods providing the needed 
transformations (e.g. $cos(\theta)$ and $\phi$ instead of $p_x, p_y, p_z$) could be 
added to the existing interface by applying decorator classes to the minimal interface.

\subsection{Technical Realization \label{sec_techreal}}

\subsubsection{Java and C++ \label{sec_javacpp}}

The user interface for Java and C++ is defined with the AID~\cite{ref_aid} 
(Abstract Interface Definition) tool from freehep.org. It allows to define the interface 
in a Java like syntax with C++ extensions and automatically creates files with Java 
interfaces and headers with pure abstract base classes for C++. AID keeps the comments 
from the original files, thus allowing to use standard tools for documenting the source 
code and the user interface. We use {\em javadoc}~\cite{ref_javadoc} and 
{\em doxygen}~\cite{ref_doxygen}\footnote{{\em doxygen} allows to use the {\em javadoc} 
syntax} for the Java and C++ implementation respectively. 
The current API documentation can be found on the LCIO homepage~\cite{ref_lcio}.

Starting from a common definition of the interface for the two languages, one could think
of having a common implementation underneath in either Java or C++. Frameworks exist 
that allow calling of either language from within the other, e.g. {\em JNI} to call 
objects coded in C++ from Java. The advantage for the developers clearly is the need to
develop and maintain only one implementation. But this comes at the price of having to 
install (and maintain) additional libraries for the user community of the other 
language. In particular one of the great advantages of Java is the machine independence 
of {\em pure Java} which would be lost when using {\em JNI}. 
This is why we decided to have two completely separate implementations in either 
language - commonality of the API is guaranteed by the use of AID and compatibility 
of the resulting files has to be assured by (extensive) testing.

\subsubsection{Fortran \label{sec_fortran}}

In order to support {\em legacy software}, e.g. the Brahms~\cite{ref_brahms} 
reconstruction program, a Fortran interface for LCIO is a requirement. Another reason for
having a Fortran interface is the fact that a lot of physicists have not yet made 
the paradigm switch towards OO-languages. So to benefit from their knowledge one has to
provide access to simulated data via Fortran.
Almost all platforms that are used in High Energy Physics have a C++ development 
environment installed. Thus by providing suitable wrapper functions it is possible 
to call the C++ implementation from a Fortran program without the need of having to 
install additional libraries.

Fortran is inherently a non-OO-like language, whereas the the API of LCIO is. 
In order to mimic the OO-calling sequences in Fortran we introduce one wrapper 
function for each objects' member function. These wrapper functions have an additional
integer argument that is converted to the C++ pointer to the corresponding 
object. By using {\tt INTEGER*8} for these integers we avoid possible problems on
64~bit architectures (as by specification there is no guarantee that C++ pointers
and Fortran {\tt INTEGER}s are the same size). 
Two additional functions are needed for every object: one for creating and one for
deleting the object. The wrapper functions are implemented using the well known 
{\tt cfortran.h}~\cite{ref_cfortran} tool for interfacing C++ and Fortran.
By applying this scheme Fortran users have to write code that basically has a one to 
one correspondence to the equivalent C++ code. This will facilitate the switching to 
Java or C++ in the future.

\subsection{Data format \label{sec_sio}}

As a first concrete data format for LCIO we chose to use SIO (Serial Input Output).
SIO has been developed at SLAC and has been used successfully in the {\em hep.lcd}
framework. It is a serial data format that is based on XDR and thus machine 
independent. While being a sequential format it still offers some OO-features, in 
particular it allows to store and retrieve references and pointers within one record.
In addition SIO includes on the fly data compression using zlib.
SIO does not provide any direct access mechanism in itself. If at some point direct 
access becomes a requirement for LCIO it has to be either incorporated into SIO or 
the underlying persistency format has to be changed to a more elaborated IO-system. 
%As mentioned in section~\ref{sec_design} this will be feasible transparent to existing 
%user code.

\section{Summary and Outlook} 
We presented LCIO, a persistency framework for linear collider simulation studies. 
LCIO defines an extensible data model for ongoing simulation studies, provides the users
with a Java, C++ and Fortran interface and uses SIO as a first implementation format.
The {\em hep.lcd} framework in the US and the European framework - based on  
{\em Mokka} and {\em Brahms} - incorporate LCIO as their persistency scheme.
The lightweight and open design of LCIO will allow to quickly adapt to future 
requirements as they arise.
Agreeing on a common persistency format could be the first step towards a common software
framework for a future linear collider. We invite all groups working in this field to join
us in using LCIO in order to avoid unnecessary duplication of effort wherever possible.


\end{document}